\documentclass[aps,pra,amsmath,amssymb,letterpaper,twocolumn]{revtex4-1}

\usepackage{hyperref}
\usepackage{graphicx}
\usepackage{color}

\begin{document}

\title{On a dual representation of the Goldstone manifold}

\author{Carlos A. Jim\'enez-Hoyos}
\email{cjimenezhoyo@wesleyan.edu}
%% \phone{+1 832 5499711}
\affiliation{Department of Chemistry, Wesleyan University, Middletown,
  CT, 06459}

\author{Rayner R. Rodr\'iguez-Guzm\'an}
\affiliation{Physics Department, Kuwait University, 13060, Kuwait}

\author{Thomas M. Henderson}
\affiliation{Department of Chemistry, Rice University, Houston,
  TX, 77005}
\affiliation{Department of Physics and Astronomy, Rice University,
  Houston, TX, 77005}

\author{Gustavo E. Scuseria}
\affiliation{Department of Chemistry, Rice University, Houston,
  TX, 77005}
\affiliation{Department of Physics and Astronomy, Rice University,
  Houston, TX, 77005}

\date{\today}

\begin{abstract}
An intrinsic wavefunction with a broken continuous symmetry can be
rotated with no energy penalty leading to an infinite set of
degenerate states known as a Goldstone manifold. In this work, we show
that a dual representation of such manifold exists that is sampled by
an infinite set of non-degenerate states. A proof that both
representations are equivalent is provided. From the work of Peierls
and Yoccoz (Proc. Phys. Soc. A {\bf 70}, 381 (1957)), it is known that
collective states with good symmetries can be obtained from the
Goldstone manifold using a generator coordinate trial wavefunction. We
show that an analogous generator coordinate can be used in the dual
representation; we provide numerical evidence using an intrinsic
wavefunction with particle number symmetry-breaking for the electronic
structure of the Be atom and one with $\hat{S}^z$ symmetry-breaking
for a H$_5$ ring. We discuss how the dual representation can be used
to evaluate expectation values of symmetry-projected states when the
norm $|\langle \Phi | \hat{P}^q | \Phi \rangle|$ becomes very small.
\end{abstract}

\maketitle

\section{Introduction}

In finite fermion systems, exact solutions to the Schr\"odinger
equation can be labelled by quantum numbers associated with the
symmetries of the Hamiltonian. Approximate solutions may break some of
those symmetries in order to develop some of the physical correlations
in the system within the constraint imposed by the form of the trial
wavefunction. As a typical example, the $U(1)$ particle-number
symmetry is spontaneously broken in the Hartree--Fock--Bogoliubov
(HFB) framework in order to account for pairing correlations
\cite{ring_schuck}. When continuous symmetries are broken, Goldstone's
theorem establishes that the trial state can be rotated with no energy
penalty (see, e.g., Ref. \onlinecite{cui2013}). The infinite set of
all degenerate, gauge-rotated states constitutes the Goldstone
manifold (GM).

In finite systems, symmetry breaking in approximate solutions is
artificial. In fact, symmetry-broken solutions can be seen as {\em
  intrinsic} or {\em deformed} states from which exact symmetries
should still be restored. It is possible to obtain symmetry-adapted
states from the intrinsic solutions using symmetry-projection
operators. Commonly used projection techniques exploit the degeneracy
within the GM in order to restore those symmetries broken in
mean-field trial states (see, e.g.,
Refs. \onlinecite{scuseria2011,jimenez2012,rodriguez2012}), as first
proposed by Peierls and Yoccoz \cite{peierls1957}.

When some of us studied the stability matrix of symmetry-projected
mean-field states \cite{thesis} we realized that, for each continuous
broken symmetry, there are two directions in which the intrinsic
wavefunction can be deformed such that the resulting
symmetry-projected state is actually equivalent. While one of the
directions was expected, as it points along the GM, the other
direction came as a surprise.

The redundancy of symmetry-projected states has been observed in
different contexts. For instance, in our own work we described how the
inclusion of a chemical potential during the optimization of trial
number-projected HFB states in a variation-after-projection framework
did not change the final result \cite{scuseria2011}. In related work,
Jensen et al. \cite{jensen1982} described an optimization method for
the antisymmetrized geminal power (AGP) wavefunction (equivalent to
number-projected HFB). The authors noted that the AGP energy is the
ratio of two homogeneous functions of the same degree and, as such,
the parameters in the AGP wavefunction are non-unique. Some of us
recently pointed out \cite{khamoshi2019} that in fact all of the
reduced density matrices of the AGP wavefunction are invariant with
respect to a multiplicative factor rescaling the geminal
parameters. Rescaling the geminal parameters is equivalent to
modifying the underlying intrinsic determinant from a number-projected
HFB point of view but this only results in a change of normalization
of the resulting AGP wavefunction (see Sec. \ref{sec:nproj} for a more
detailed discussion).

The aim of this work is to unambiguously identify the source of this
redundancy in symmetry-projected states. Namely, we provide a proof
that an equivalent representation of the GM exists. This is generated
by a set of {\em non-degenerate} states, generated from the parent
broken symmetry state using a rotation involving a purely imaginary
angle.

The rest of the manuscript is organized as follows. In
Sec. \ref{sec:sb} we introduce the GM associated with symmetry-broken
trial wavefunctions. In Sec. \ref{sec:gcm} we provide a short
introduction to the Generator Coordinate Method (GCM), which we use in
Sec. \ref{sec:gcm2} to extract symmetry-adapted states from the GM. In
Sec. \ref{sec:dp} we provide a short discussion of how the double
projection method relates to the dual representation of the GM. In
Sec. \ref{sec:results} we provide the dual representations of the GMs
associated with particle-number symmetry breaking and $\hat{S}^z$
symmetry breaking. We also discuss the shape of the eigenstates of the
Hamiltonian obtained from the dual representation of the GM. In
Sec. \ref{sec:conclusions} we make some concluding remarks, including
how the dual representation can be used, in certain cases, to improve
the numerical precision in the evaluation of matrix elements between
symmetry-projected states.

\section{Theory}

\subsection{Symmetry Breaking and the Goldstone Manifold}
\label{sec:sb}

Continuous symmetries of the Hamiltonian can be either Abelian or
non-Abelian, depending on the number of generators involved and their
commutation properties. For the sake of simplicity, we will limit
ourselves in this work to Abelian groups, that is, $U(1)$
symmetries. Nonetheless, the dual representation described below also
exists in non-Abelian symmetries.

Consider an operator $\hat{Q}$ associated with a continuous symmetry
of the Hamiltonian such that $[\hat{Q},\hat{H}]=0$. Note that
$\hat{Q}$ must be Hermitian in order to be associated with an
observable. Exact eigenstates of $\hat{H}$ can always be chosen as
eigenstates of $\hat{Q}$. Approximations to eigenstates need not,
however, preserve the symmetry. If a trial state $|\Phi \rangle$ is
not an eigenfunction of $\hat{Q}$, then the states
\begin{equation}
  |\Phi_\theta \rangle = \exp(i\theta \hat{Q}) \, |\Phi \rangle
  \label{eq:expiQ}
\end{equation}
constitute the GM. We label this representation as {\em direct} in the
rest of this manuscript. It is trivial to show that the unitary
operator $\exp(i\theta \hat{Q})$ preserves the norm and Hamiltonian
expectation values
\begin{align}
  \langle \Phi_\theta | \Phi_\theta \rangle &= \, \langle \Phi | \Phi
  \rangle, \\
  \langle \Phi_\theta | \hat{H} | \Phi_\theta \rangle &= \, \langle
  \Phi | \hat{H} | \Phi \rangle.
\end{align}
That is, all states in the direct representation are degenerate. A
rotation in the GM can therefore be performed with no energy penalty.

The central premise of this manuscript is that there exists a {\em
  dual} representation of the same GM. This is generated by the set of
states
\begin{equation}
  |\Phi_{\vartheta} \rangle = \mathcal{N} \, \exp(\vartheta \hat{Q})
  \, |\Phi \rangle,
  \label{eq:expQ}
\end{equation}
where $\mathcal{N}$ is a normalization factor.  (We use, for
convenience, different symbols for rotations in the direct ($\theta$)
and dual $(\vartheta)$ representations.) Note that $\exp(\vartheta
\hat{Q})$ is not unitary and therefore the norm and Hamiltonian
expectation values are not preserved. We refer the reader to App.
\ref{sec:eqGM} for a proof of the equivalence of the manifolds.

\subsection{Generator Coordinate Method (GCM)}
\label{sec:gcm}

In this section we provide a short introduction to the GCM, with only
the necessary elements to follow the discussion below. For more
details about the GCM, we refer the reader to
Refs. \onlinecite{ring_schuck,reinhard1987}. In the GCM, a variational
ansatz for the wavefunction is written as
\begin{equation}
  |\psi \rangle = \int da \, |\phi_a \rangle \, f(a).
\end{equation}
The wavefunction $|\psi \rangle$ is written as a superposition, with a
weight function $f(a)$, of the intrinsic wavefunctions $|\phi_a
\rangle$ along the deformation parameter $a$. Most commonly, $|\phi_a
\rangle$ are chosen as mean-field states, but this need not always be
the case.

The function $f(a)$ can be determined by the variational principle
leading to the Griffin-Hill-Wheeler (GHW) equation:
\begin{equation}
  \int da' \, [ \mathcal{H}(a,a') - E \, \mathcal{S}(a,a') ] \, f(a')
  = 0,
\end{equation}
with $\mathcal{H}(a,a') = \langle \phi_a | \hat{H} | \phi_a' \rangle$
and $\mathcal{S}(a,a') = \langle \phi_a | \phi_a' \rangle$. The state
$|\psi \rangle$ can be normalized as
\begin{equation}
  \langle \psi | \psi \rangle = 1 = \int da \, \int da' \, f^\ast(a)
  \, \mathcal{S}(a,a') \, f(a').
\end{equation}

Because of this normalization choice, the function $|f(a)|^2$ cannot
be associated with a probability distribution. On the other hand, it
is possible to construct
\begin{equation}
  g(a) = \int da' \, \mathcal{S}^{1/2}(a,a') \, f(a').
\end{equation}
It follows that $|g(a)|^2$ can actually be associated with a
probability distribution:
\begin{equation}
  \int da \, |g(a)|^2 = 1
\end{equation}
Here, $\mathcal{S}^{1/2}$, the operational square root of the norm
kernel, can be formally defined through
\begin{equation}
  \mathcal{S}(a,a') = \int da'' \, \mathcal{S}^{1/2}(a,a'') \,
  \mathcal{S}^{1/2}(a'',a')
\end{equation}

It will also prove useful to introduce the function
\begin{equation}
  h(a) = \int da' \, \mathcal{S}(a,a') \, f(a') = \langle \phi_a |
  \psi \rangle.
\end{equation}
Here, $h(a)$ gives the projection of $|\psi \rangle$ onto the
component $|\phi_a \rangle$. As described in the next section, a GCM
ansatz can be used along the GM to yield symmetry-adapted states.

\subsection{GCM on the Goldstone manifold}
\label{sec:gcm2}

Peierls and Yoccoz \cite{peierls1957} proposed the use of a GCM ansatz
among the states in the GM. (See also
Refs. \onlinecite{wong1975,ring_schuck} for a more detailed
presentation.)  Namely, they proposed a variational optimization of an
ansatz of the form
\begin{equation}
  |\Psi \rangle = \int d\theta \, |\Phi_\theta \rangle \,
  \tilde{f}(\theta),
  \label{eq:gcmth}
\end{equation}
with $|\Phi_\theta \rangle$ given by Eq. \ref{eq:expiQ}.

Note that all the states $|\Phi_\theta \rangle$ are spanned by the
same set of symmetry-adapted states (see App. \ref{sec:eqGM}). A
representation of the Hamiltonian among such symmetry-adapted states
is necessarily diagonal given that $\hat{Q}$ is a symmetry of
$\hat{H}$. It follows that the eigenstates of the Hamiltonian among
the overcomplete set $\{ \Phi_\theta \}$ correspond to the
symmetry-adapted states. In other words, symmetries can be restored by
a wavefunction of the form of Eq. \ref{eq:gcmth}. We can therefore
rewrite the ansatz as
\begin{equation}
  |\Psi^q \rangle = \int d\theta \, |\Phi_\theta \rangle \,
  \tilde{f}^q(\theta),
\end{equation}
which explicitly indicates that the weight function
$\tilde{f}^q(\theta)$ depends parametrically on the quantum number $q$
to yield the symmetry-adapted state $|\Psi^q \rangle$.

The same justification can be established by a GCM ansatz on the dual
GM. Namely, the ansatz
\begin{equation}
  |\Psi^q \rangle = \int d\vartheta \, |\Phi_{\vartheta} \rangle \,
  f^q(\vartheta),
  \label{eq:gcmvth}
\end{equation}
with $|\Phi_{\vartheta} \rangle$ given by Eq. \ref{eq:expQ}, can also
be used and will yield the same set of symmetry-adapted states. The
amplitudes $f^q(\vartheta)$ can be obtained from a solution to the
corresponding GHW equation.

In the direct representation, it is possible to deduce the form of the
amplitudes in the GCM ansatz of Eq. \ref{eq:gcmth} without solving the
corresponding GHW equation. For instance, in the case of a $U(1)$
symmetry, the GCM amplitudes are given by (see App. \ref{sec:kernel})
\begin{equation}
  \tilde{f}^q (\theta) = \frac{1}{2\pi} e^{-i\theta q}.
\end{equation}
This can be used to deduce the form of the projection operator for an
Abelian symmetry as
\begin{equation}
  \hat{P}^q = \frac{1}{2\pi} \int d\theta \, \exp(i\theta(\hat{Q} -
  q)).
\end{equation}

In the dual representation one cannot generally deduce the shape of
the weight functions $f^q(\vartheta)$ a priori. These have to be
obtained by a numerical solution to the GHW equations.  We note that,
in practical calculations, one can use a discretized sampling of the
domain $\vartheta$. In particular, it can be easily shown that the
exact Hamiltonian eigenvalues can be obtained as long as the number of
grid points sampled is sufficient to obtain all quantum numbers. For
$\hat{N}$ projection, this is equal to $M+1$, where $M$ is the number
of orbitals available.

\subsection{Comparison with Double Projection}
\label{sec:dp}

In Ref. \onlinecite{peierls1962}, Peierls and Thouless introduced the
double projection (DP) method or double GCM (DGCM). In DGCM, the
deformation parameter is generalized from a real to a complex
one. This extension allowed the authors to obtain the correct kinetic
energy associated with translational motion in nuclei.

In Sec. \ref{sec:sb} we established that the manifold generated by a
real or a purely imaginary deformation parameter are
equivalent. Therefore, there is no additional variational flexibility
gained by letting the generator coordinate become complex.

This, however, does not apply to DGCM. In particular, the idea of
Peierls and Thouless was to define a manifold of intrinsic states
$|\Phi_x \rangle$ where these are obtained by minimization of $\langle
\hat{H} \rangle$ subject to the constraint $\langle \hat{Q} \rangle =
x$. (In DGCM, the authors additionally restore the symmetry $\hat{Q}$,
for each $|\Phi_x \rangle$, using the direct representation of the
GM.) The manifold $\{ \Phi_x \}$ is not equivalent to the dual
representation of the GM, where the states $|\Phi_{\vartheta} \rangle$
are generated by a simple rotation of $|\Phi \rangle$. In particular,
it will not be generally true that $\exp(\vartheta \hat{Q}) |\Phi
\rangle$ is a minimizer of $\langle \hat{H} \rangle$, even though it
does satisfy the constraint imposed.

\section{Results and Discussion}
\label{sec:results}

\subsection{$\hat{N}$ projection on Be}
\label{sec:nproj}

Due to the repulsive nature of electron-electron interactions,
spontaneous symmetry breaking of particle number does not occur in
mean-field solutions to the electronic Schr\"odinger equation in
molecular and atomic systems \cite{bach1994}. An intrinsically
deformed state can still be obtained when the wavefunction is
optimized in the presence of a projection operator ({\em i.e.}, a
variation-after-projection approach). We work with a number-projected
HFB solution to the neutral Be atom ($n=4$) using a standard cc-pVDZ
basis set \cite{dunning1989}.

The intrinsic wavefunction used corresponds to an HFB state with
singlet pairing and can be written, in the natural orbital basis as
\begin{equation}
  |\Phi \rangle = \prod_k \left( 1 + \frac{v_k}{u_k}
  a^\dagger_{k\uparrow} a^\dagger_{k\downarrow} \right) |-
  \rangle, \label{eq:hfb}
\end{equation}
where $|- \rangle$ is the bare vacuum.  We have optimized the
wavefunction under the contraint $\langle \Phi | \hat{N} | \Phi
\rangle / \langle \Phi | \Phi \rangle = 4$.

\begin{figure*}
  \includegraphics[width=8cm]{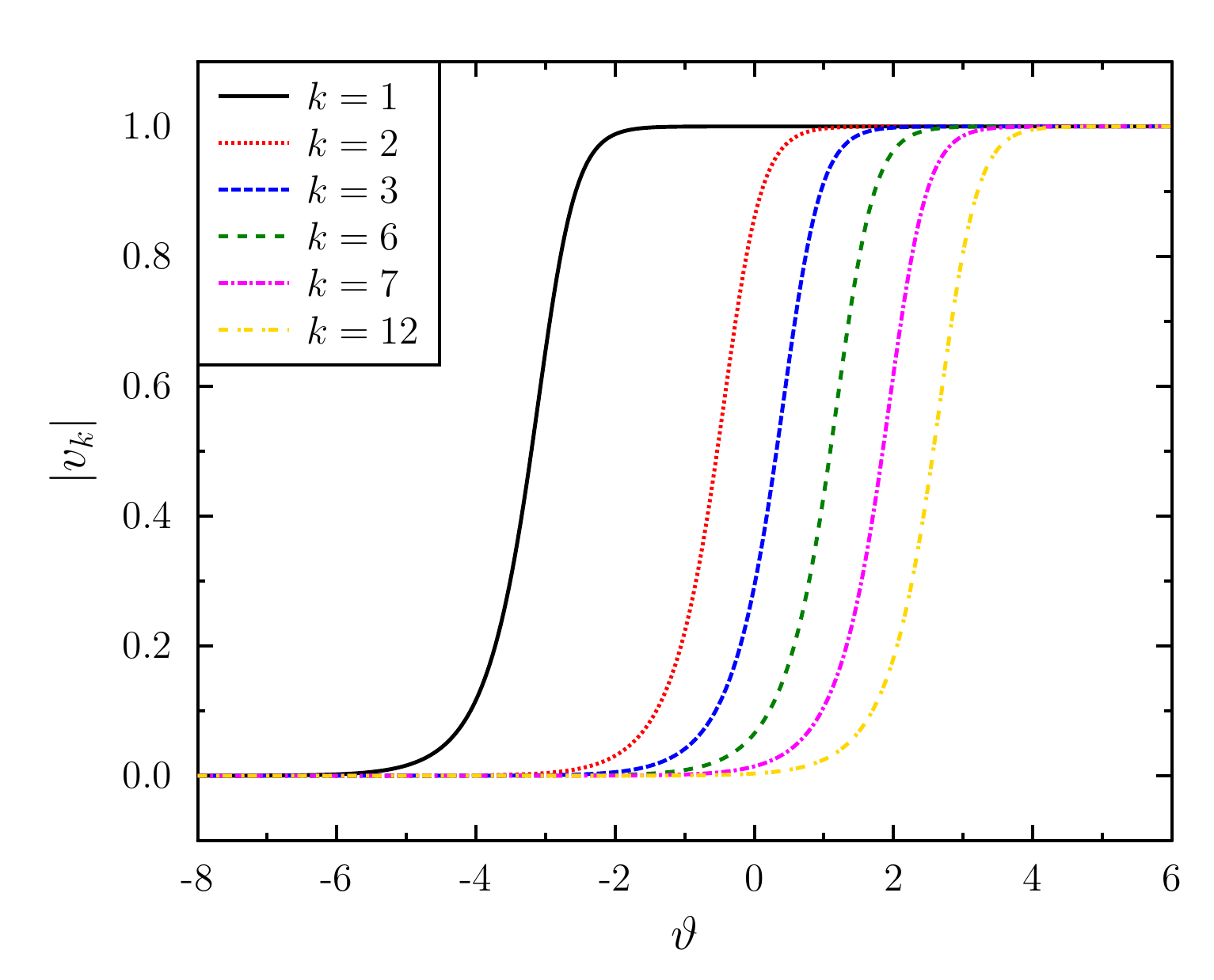}
  \hspace{0.2cm}
  \includegraphics[width=8cm]{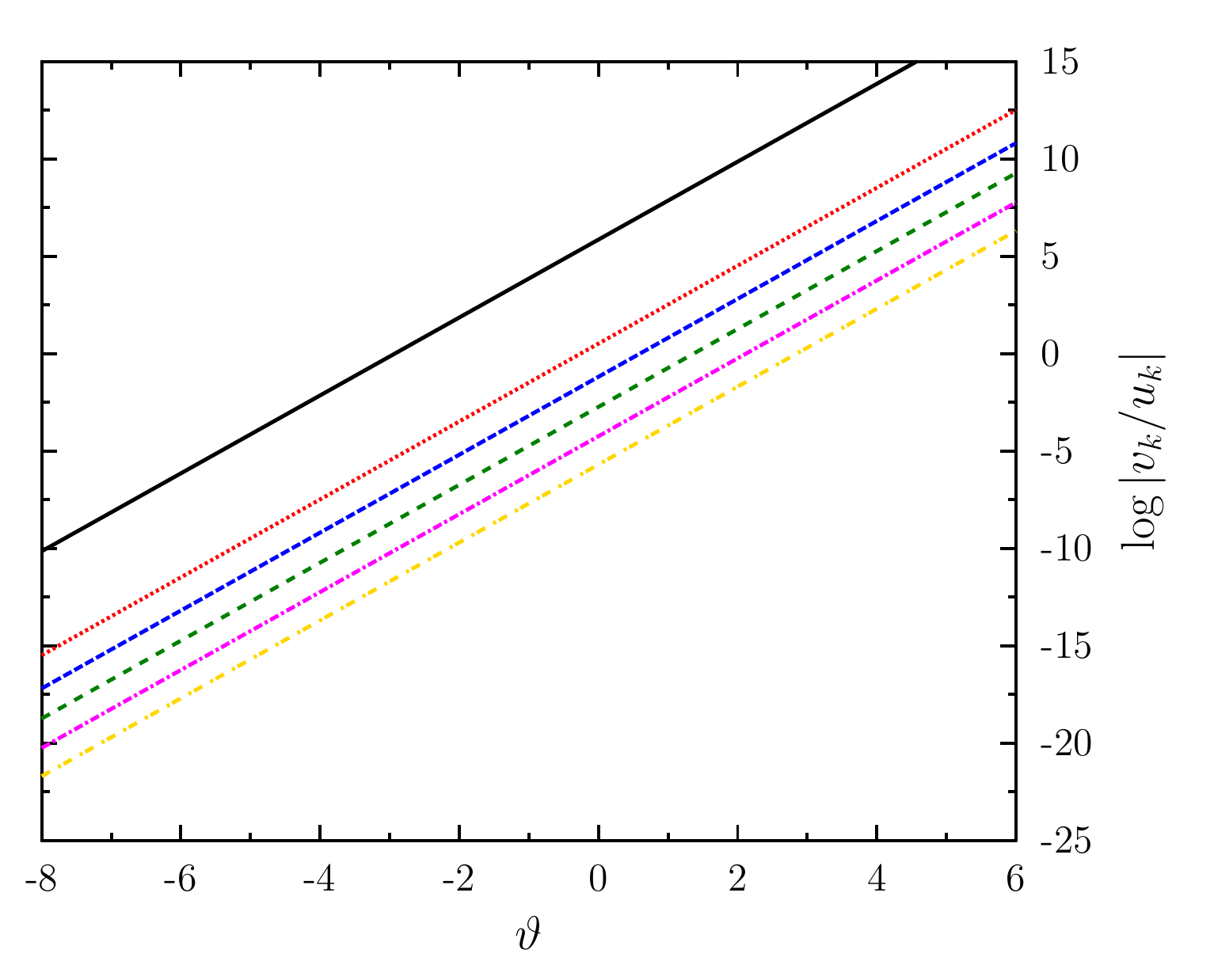}
  \caption{(Left) Coefficients $|v_k|$ in the intrinsic wavefunction
    along the deformation parameter $\vartheta$. Note that some
    levels, corresponding to $p$ or $d$ orbitals, are multiply
    degenerate: $v_3 = v_4 = v_5$, $v_7 = \cdots = v_{11}$, and
    $v_{12} = v_{13} = v_{14}$. (Right) Ratios $|v_k/u_k|$ in the
    intrinsic wavefunction. \label{fig:be_avp}}
\end{figure*}

The dual representation of the GM corresponds to the set of states
\begin{equation}
  |\Phi_{\vartheta} \rangle = \mathcal{N} \, \exp(\vartheta \hat{N}) \,
  |\Phi \rangle,
\end{equation}
where $\mathcal{N}$ is a normalization factor \footnote{We work with
  HFB states that follow the usual normalization $|v_k|^2 + |u_k|^2 =
  1$.}. The set $\{ \Phi_{\vartheta} \}$ is composed of HFB states of
the same form as that of Eq. \ref{eq:hfb}, expanded in the same set of
natural orbitals, but with parameters ${v_k}$ that depend on the angle
$\vartheta$. This dependence is shown in Fig. \ref{fig:be_avp}.

\begin{figure*}
  \includegraphics[width=8cm]{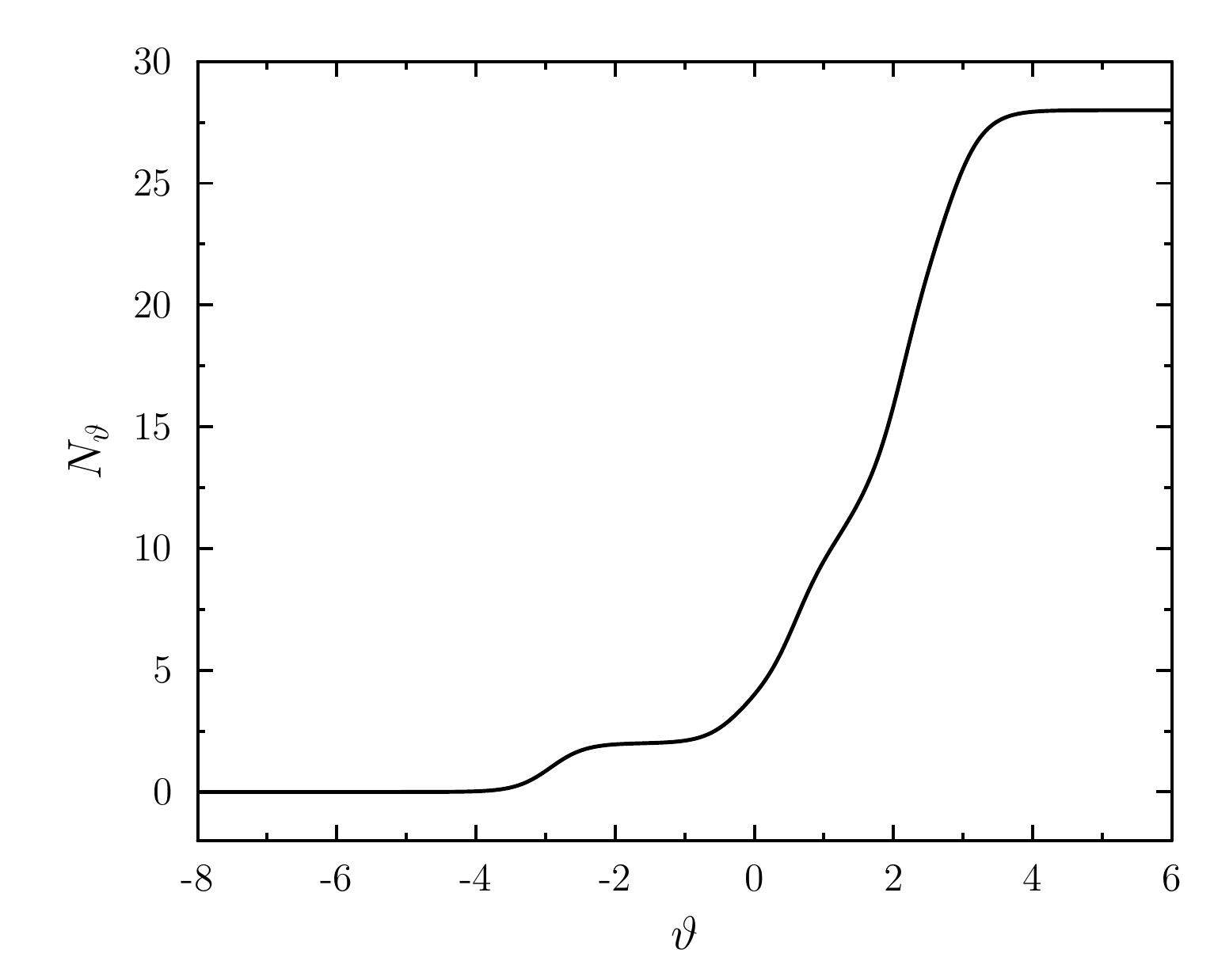}
  \hspace{0.2cm}
  \includegraphics[width=8cm]{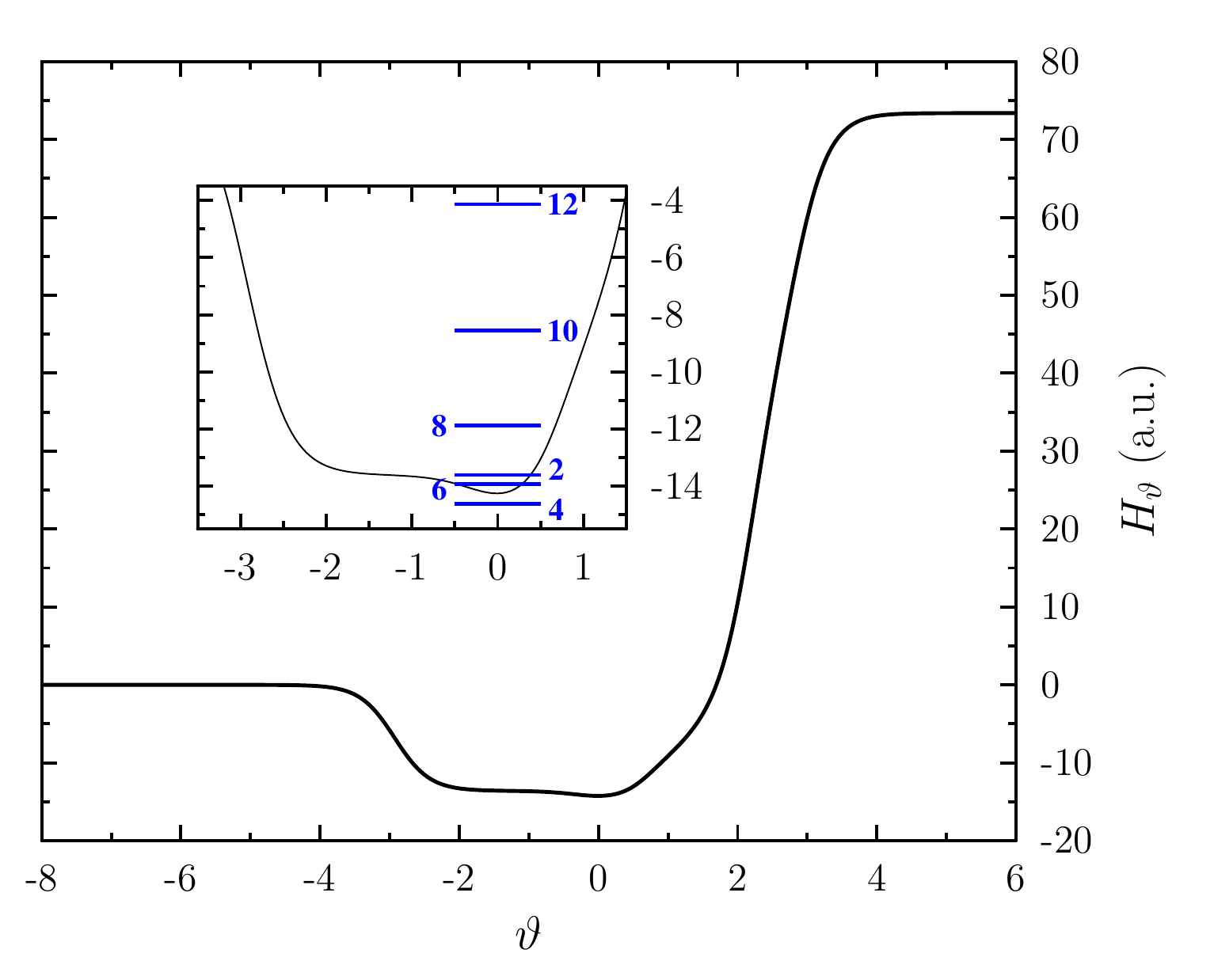}
  \caption{(Left) Expectation value of the number operator $\hat{N}$
    along the deformation parameter $\vartheta$. (Right) Expectation
    value of the Hamiltonian operator $\hat{H}$. The inset zooms into
    the low-energy region; it also shows the energies of the
    symmetry-restored states (with the value of $n$ indicated) which
    are equivalent to the eigenvalues obtained from a solution of the
    GHW equation in the dual GM. \label{fig:be_nh}}
\end{figure*}

We show in Fig. \ref{fig:be_nh} the expectation values of $\hat{N}$
and $\hat{H}$ among the states $|\Phi_{\vartheta} \rangle$, with
\begin{align}
  N_{\vartheta} &= \, \langle \Phi_{\vartheta} | \hat{N} | \Phi_{\vartheta}
  \rangle / \langle \Phi_{\vartheta} | \Phi_{\vartheta} \rangle, \\
  H_{\vartheta} &= \, \langle \Phi_{\vartheta} | \hat{H} | \Phi_{\vartheta}
  \rangle / \langle \Phi_{\vartheta} | \Phi_{\vartheta} \rangle.
\end{align}
As it is evident from the figure, expectation values are not conserved
in this dual representation. $N_{\vartheta}$ varies monotonically
between 0 and 28 (the basis set used has 14 basis
functions). $H_{\vartheta}$ has a minimum near $\vartheta = 0$, as
expected, given that $\vartheta = 0$ was defined such that
$N_{\vartheta} = 4$, corresponding to a neutral Be atom.

\begin{figure*}
  \includegraphics[width=8cm]{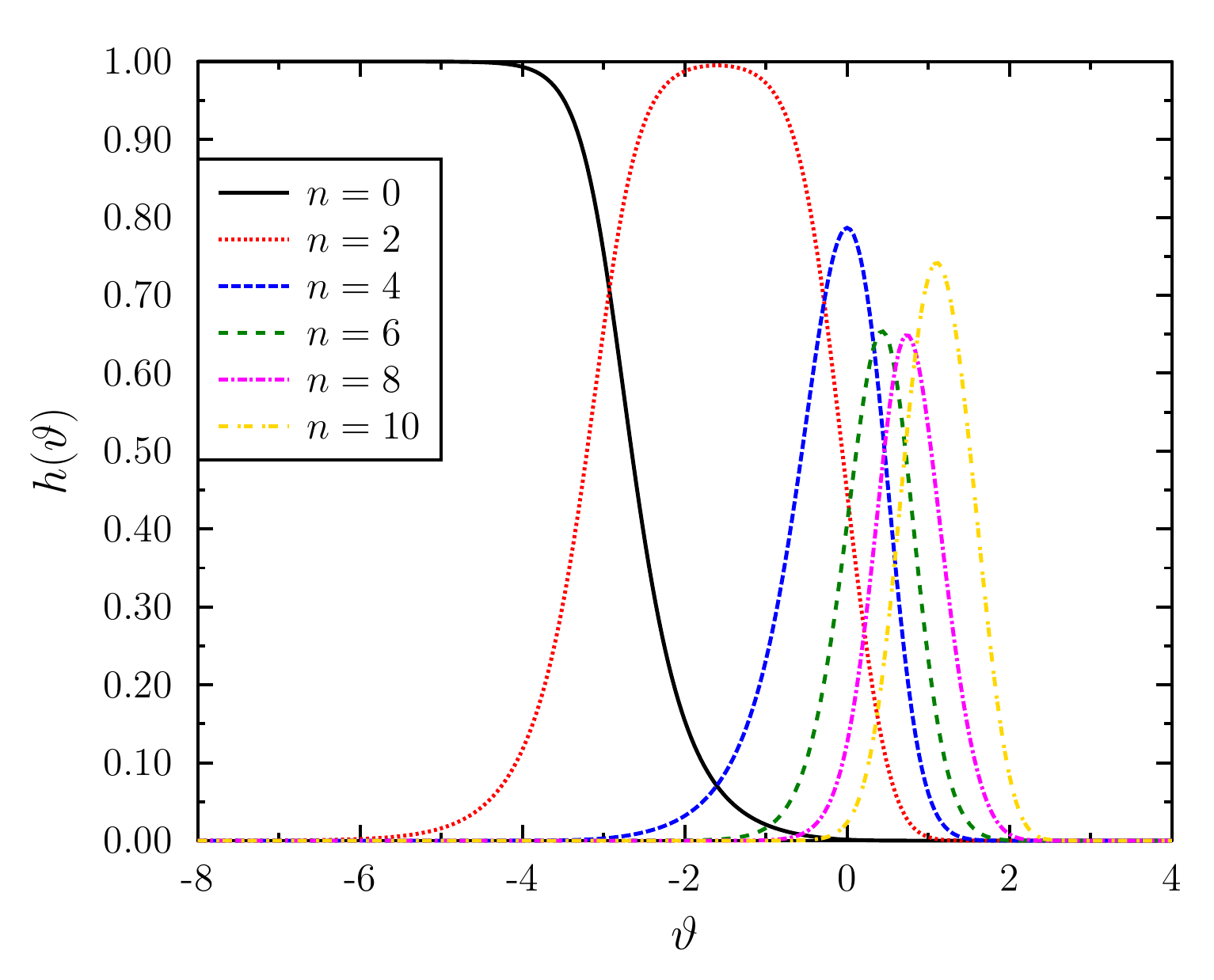}
  \hspace{0.2cm}
  \includegraphics[width=8cm]{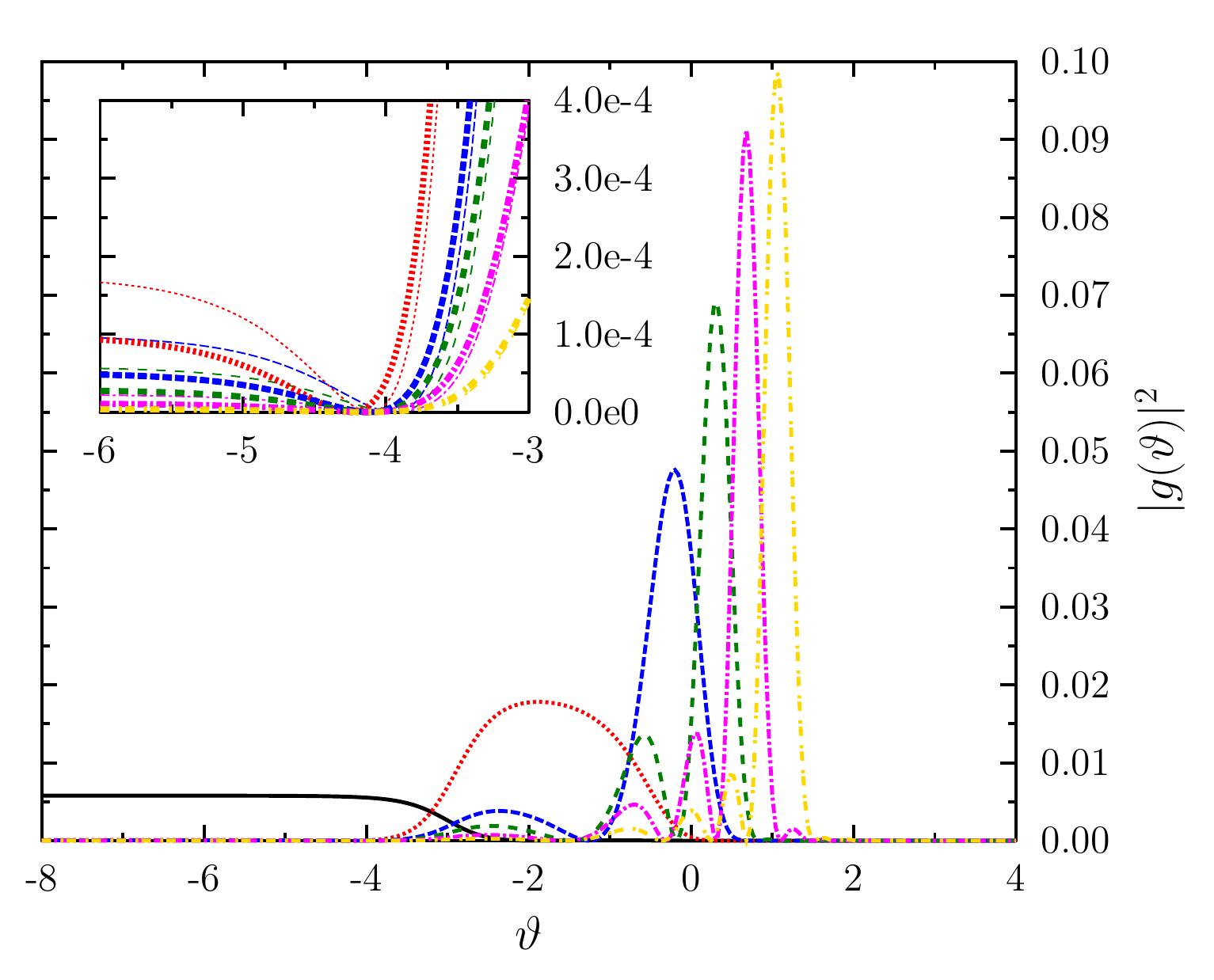}
  \caption{(Left) GCM wavefunctions $h(\vartheta)$ obtained as
    solutions to the GHW equation for $n=0,2,\ldots,10$. (Right)
    Probability distributions $|g(\vartheta)|^2$ obtained with a
    deformation domain $\vartheta \in (-10,8)$. The inset shows, in
    thin lines, the behavior for large $-\vartheta$ when the
    deformation domain is reduced to $\vartheta \in
    (-8,6)$. \label{fig:be_wfn}}
\end{figure*}

The wavefunctions obtained from the solution of the GHW equation in
the dual representation of the GM are shown in
Fig. \ref{fig:be_wfn}. On the right, we show the probability
distributions $|g(\vartheta)|^2$. Note that, in solving for the GHW
equation, we have pre-normalized each intrinsic state, such that the
norm kernel has a unit diagonal.

The form of $g(\vartheta)$ depends on the domain chosen for
$\vartheta$, as shown in the inset of Fig. \ref{fig:be_wfn}. Here, we
should clarify that the domain of $\vartheta$ can be formally
truncated to a finite one: as long as the domain samples the states of
all symmetries $q$, enlarging the domain just adds redundant
information. Using an infinite domain for $\vartheta$ renders some
states (in this case, those with $n=0$ or $n=28$) non-normalizable.
Unlike $g(\vartheta)$, the representation $h(\vartheta)$ is unique,
given that this is the projection of the eigenstates onto the manifold
$\{ \Phi_{\vartheta} \}$. This can be obtained, even without solving
the GHW equation, using the known representation of the
symmetry-adapted states in the direct manifold.

\begin{figure*}
  \includegraphics[width=8cm]{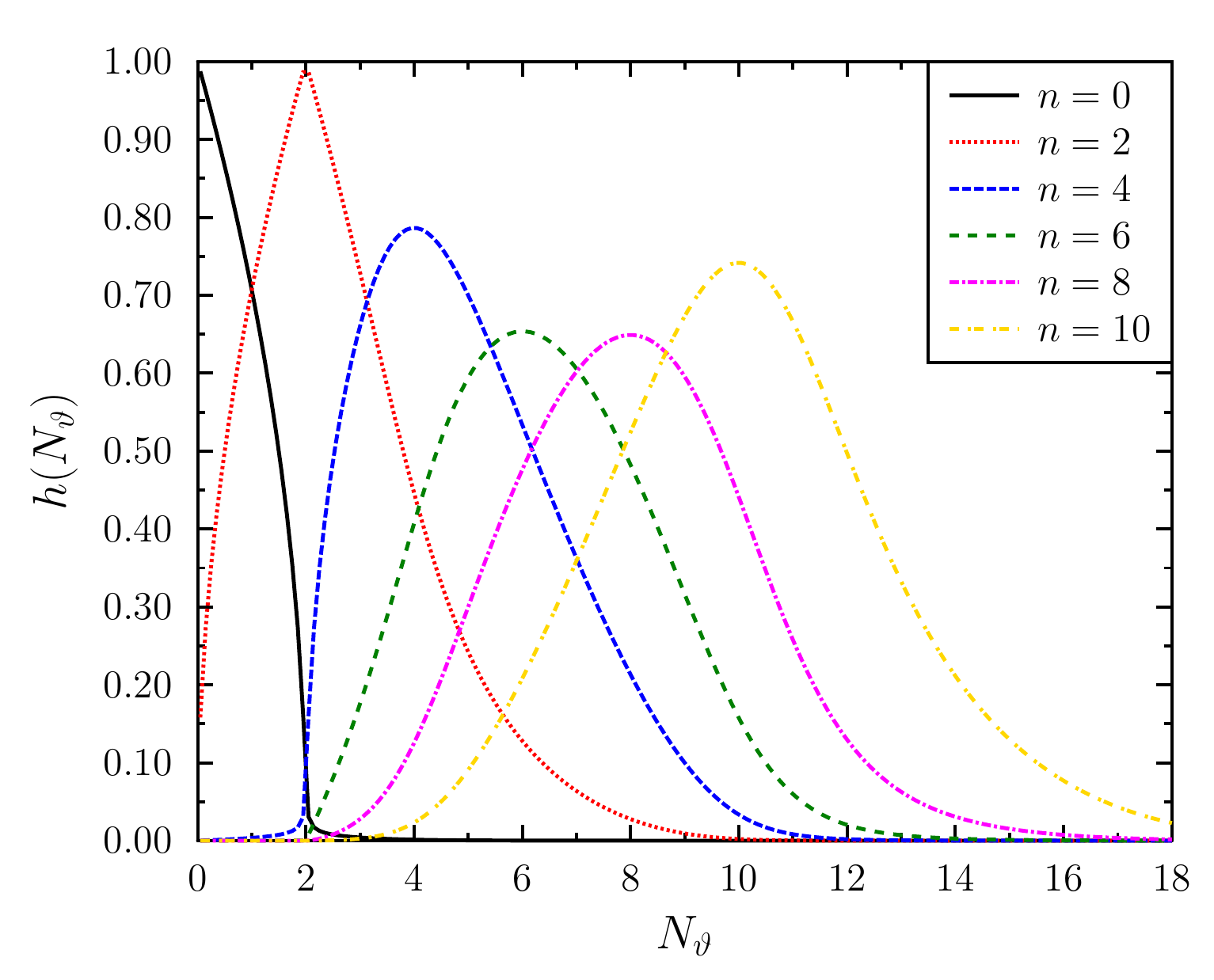}
  \hspace{0.2cm}
  \includegraphics[width=8cm]{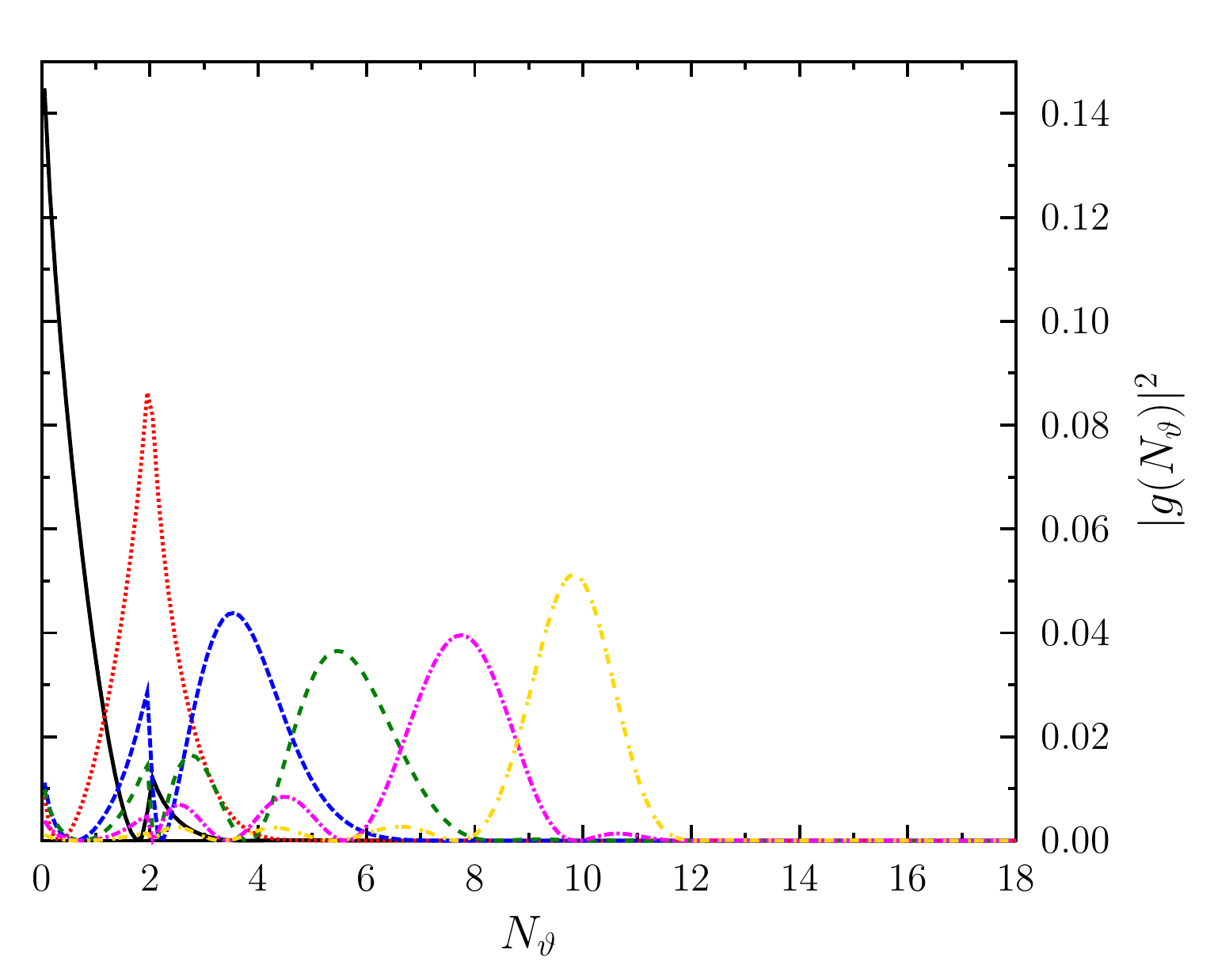}
  \caption{Same as Fig. \ref{fig:be_wfn}, but mapping the deformation
    parameter $\vartheta$ onto $N_{\vartheta}$. \label{fig:be_wfn2}}
\end{figure*}

Because $N_{\vartheta}$ varies monotonically with $\vartheta$, the
deformation parameter $\vartheta$ can be mapped onto
$N_{\vartheta}$. The resulting amplitudes, as a function of
$N_{\vartheta}$, are shown in Fig. \ref{fig:be_wfn2}. As one would
expect, $h(N_{\vartheta})$ peaks at $N_{\vartheta} = n$. For $n \geq
6$ the functions have a Gaussian-like profile, suggesting that a
numerical integration with an appropriate Gaussian quadrature can
yield a good description of such states.

\subsection{$\hat{S}^z$ projection on H$_5$ ring}

\begin{figure}
  \includegraphics[width=5cm]{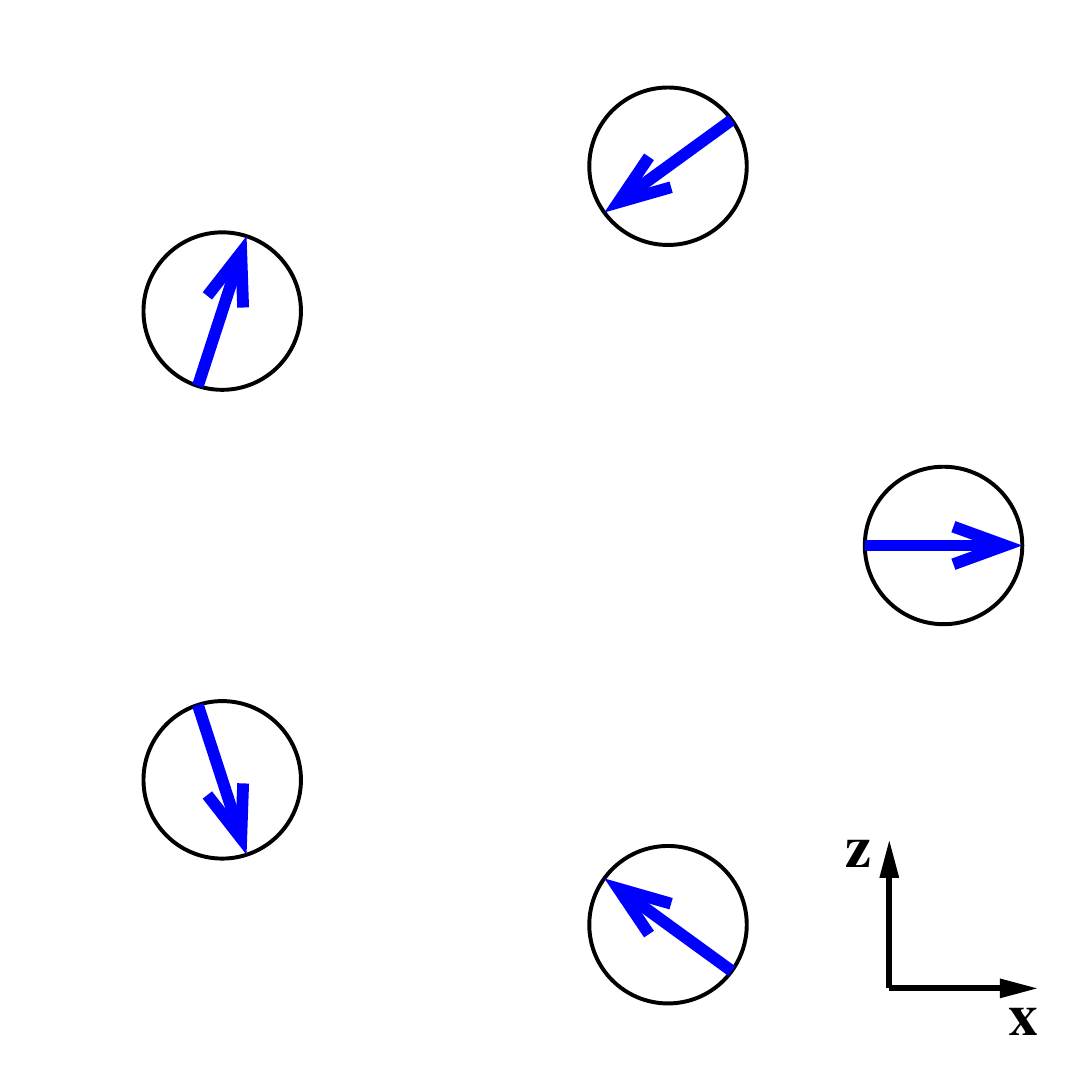}
  \caption{Coplanar spin arrangement in the HF solution for H$_5$. The
    axis system in the lower corner indicates the convention used for
    the spin axis. \label{fig:h5_spin}}
\end{figure}

We next consider a system with five hydrogen atoms equally spaced
placed around a circle such that the distance between nearest neighbor
atoms is $1.8$~a.u. A standard cc-pVDZ basis set \cite{dunning1989}
was used in the calculations (5 basis functions per H atom). The
lowest-energy Hartree--Fock (HF) solution for a ring of hydrogen atoms
produces an anti-ferromagnetic alignment of the spins; when the rings
have an odd number of atoms, this leads to spin frustration and a
generalized HF (GHF) ground state with a coplanar spin arrangement
\cite{li2015}. In the case of H$_5$, the lowest energy solution has
nearest-neighbor spins rotated by 144 degrees, as illustrated
schematically in Fig. \ref{fig:h5_spin}.

As discussed in Ref. \onlinecite{henderson2018}, a HF solution with a
coplanar but not collinear spin arrangement is not an eigenfunction of
$\hat{S}^q$, for any direction $q$. In what follows, we consider the
restoration of $\hat{S}^z$ as a symmetry from a solution with spins
oriented as shown in Fig. \ref{fig:h5_spin}. (Note that results would
be different had we chosen to restore $\hat{S}^x$ or $\hat{S}^y$. This
already suggests that the right symmetry to restore is the full spin
rotation, i.e., $\hat{S}^2$ symmetry; we proceed, nonetheless with
$\hat{S}^z$ restoration for simplicity.)

The intrinsic wavefunction used corresponds to a GHF state which can
be written in a corresponding orbital basis (with the spin-up orbitals
being different than the spin-down ones, even when they are spanned by
the same basis functions) as
\begin{equation}
  |\Phi \rangle = \prod_{k=1}^N \left( v_k
  a^\dagger_{k\uparrow} + u_k a^\dagger_{k\downarrow}
  \right) |- \rangle, \label{eq:hf}
\end{equation}
where $N$ is the number of electrons. As it should be evident from the
scheme \ref{fig:h5_spin}, the intrinsic wavefunction satisfies
$\langle \Phi | \hat{S}^z | \Phi \rangle / \langle \Phi | \Phi \rangle
= 0$.

\begin{figure*}
  \includegraphics[width=8cm]{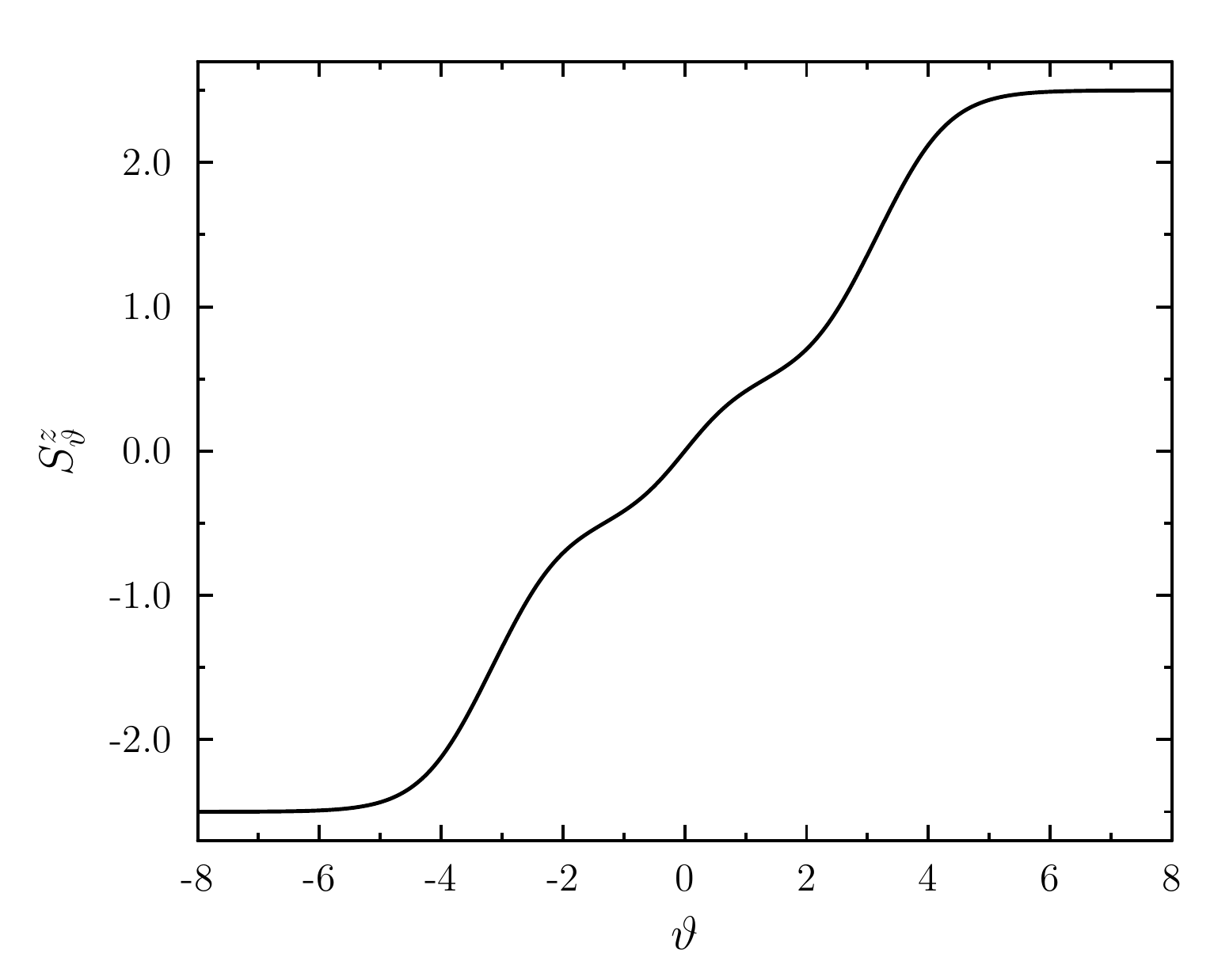}
  \hspace{0.2cm}
  \includegraphics[width=8cm]{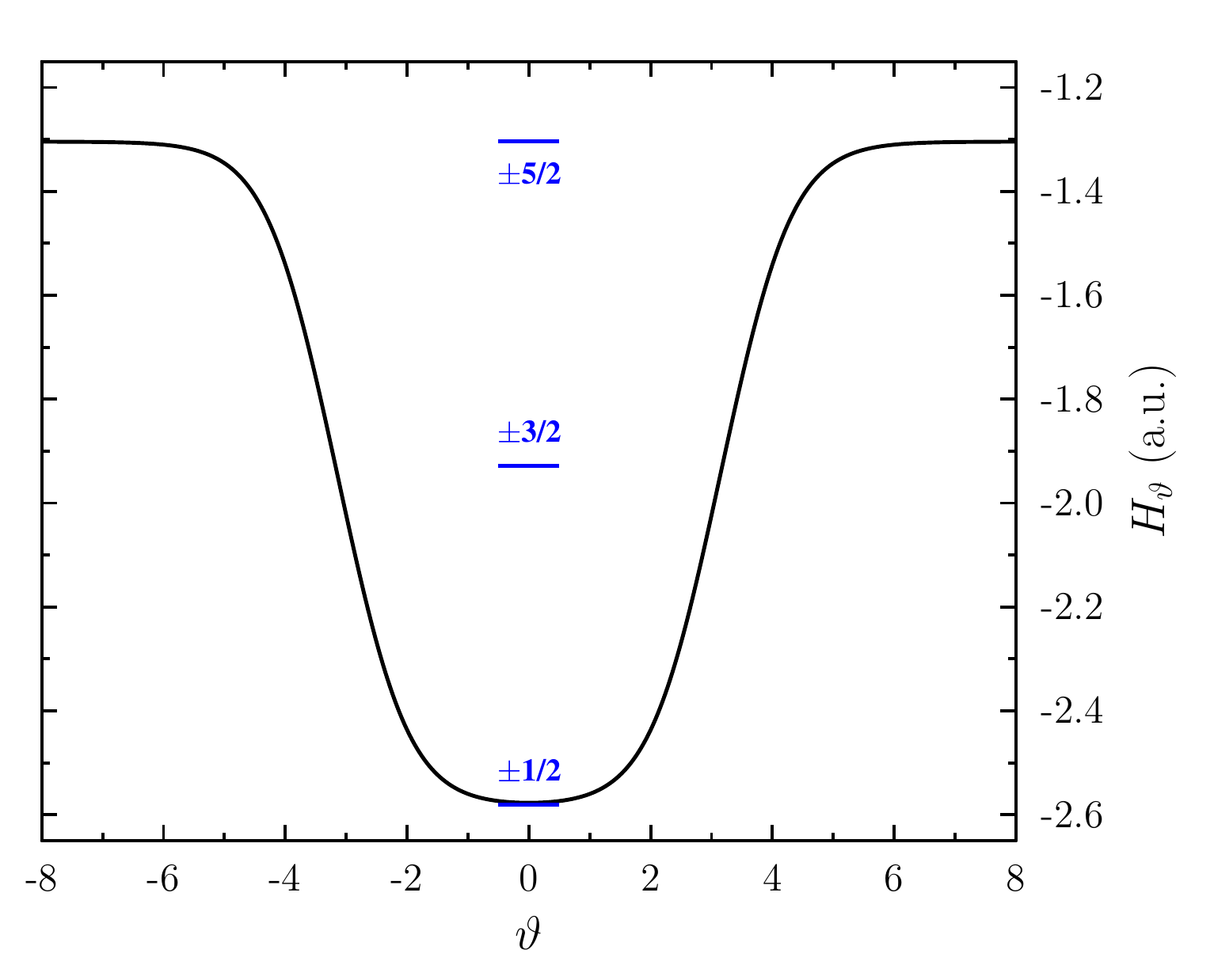}
  \caption{(Left) Expectation value of $\hat{S}^z$ along the
    deformation parameter $\vartheta$. (Right) Expectation value of
    the Hamiltonian operator $\hat{H}$. The energies of the
    symmetry-restored states are also shown (with the value of $s^z$
    indicated); states with $\pm s^z$ are exactly
    degenerate. \label{fig:h5_szh}}
\end{figure*}

The dual representation of the GM is generated in this case through
\begin{equation}
  |\Phi_{\vartheta} \rangle = \mathcal{N} \, \exp(\vartheta \hat{S}^z)
  |\, \Phi \rangle,
\end{equation}
where $\mathcal{N}$ is a normalization factor \footnote{We work with
  normalized HF states that satisfy $|v_k|^2 + |u_k|^2 =
  1$.}. Fig. \ref{fig:h5_szh} shows the dependence of the expectation
values of $\hat{H}$ and $\hat{S}^z$ on the deformation parameter
$\vartheta$. Note that in this case $H_\vartheta$ is even while
$S^z_\vartheta$ is odd about $\vartheta = 0$.

\begin{figure*}
  \includegraphics[width=8cm]{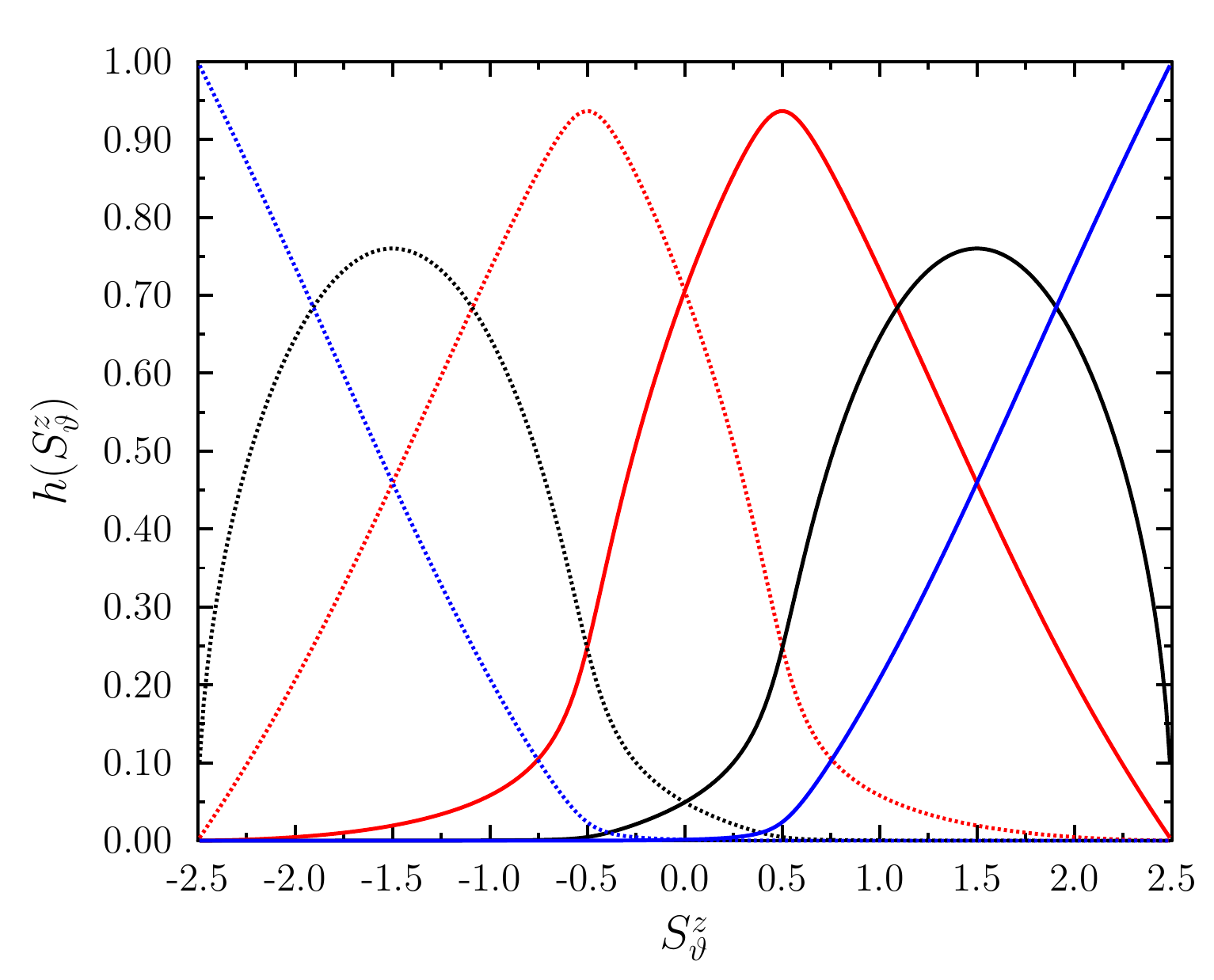}
  \hspace{0.2cm}
  \includegraphics[width=8cm]{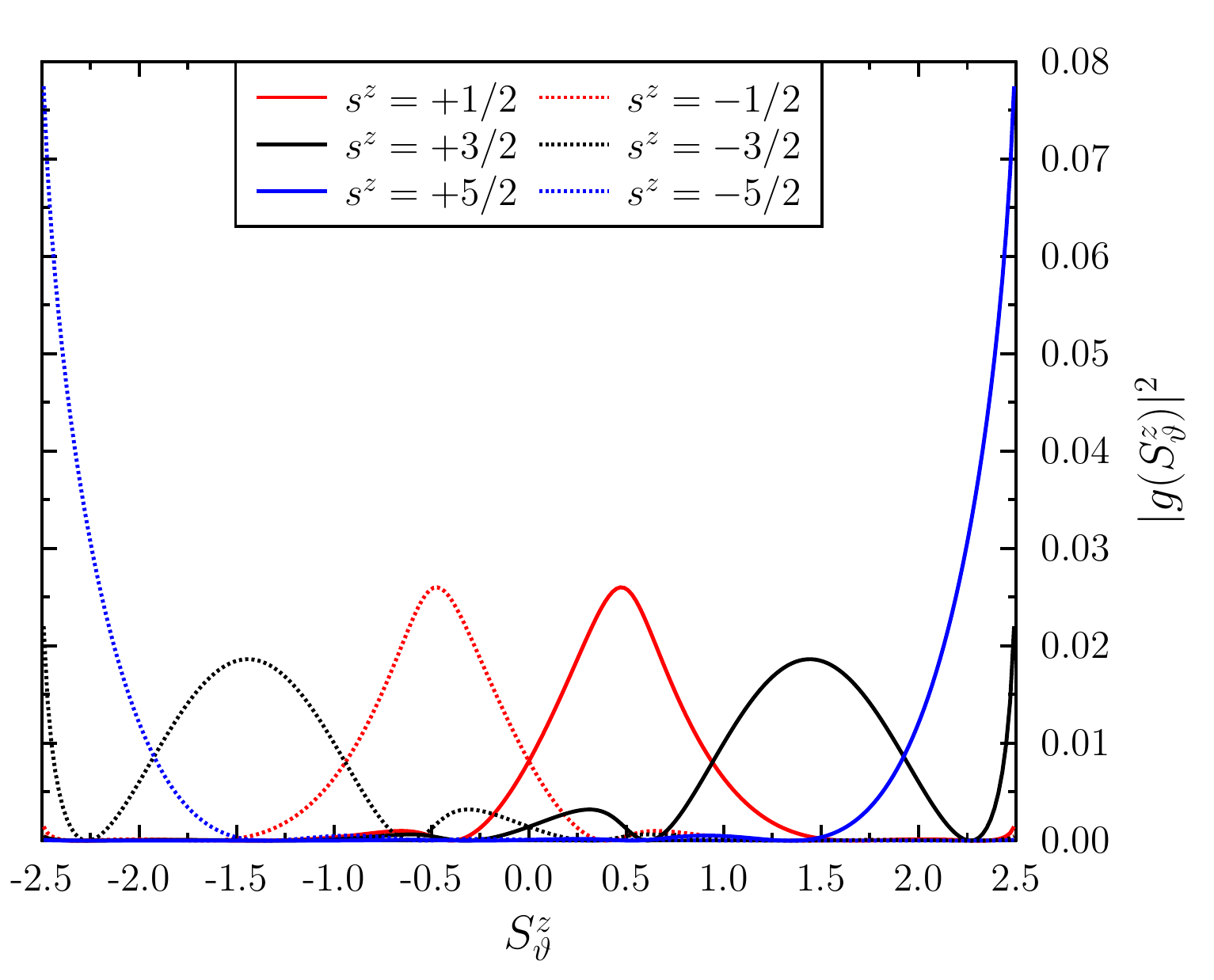}
  \caption{(Left) GCM wavefunctions $h(S^z_{\vartheta})$ obtained as
    solutions to the GHW equation where the deformation parameter
    $\vartheta$ has been mapped onto $S^z_{\vartheta}$. (Right)
    Probability distributions
    $|g(S^z_{\vartheta})|^2$. \label{fig:h5_wfn}}
\end{figure*}

The wavefunctions obtained from the solution of the GHW equation in
the dual representation of the GM are shown in
Fig. \ref{fig:h5_wfn}. In this case, the wavefunctions are only shown
with the deformation parameter $\vartheta$ mapped onto
$S^z_{\vartheta}$. The wavefunction amplitudes for $\pm s^z$ are
reflections of each other through $S^z_{\vartheta} = 0$, as one would
have expected. Note that the decomposition of $|\Phi_{\vartheta}
\rangle$ in terms of symmetry-adapted states can be extracted from the
profile of $h(S^z_{\vartheta})$ in Fig. \ref{fig:h5_wfn}.

\section{Conclusions}
\label{sec:conclusions}

In this work, we have shown that a dual representation exists for the
Goldstone manifold of symmetry-broken states. This is a general
conclusion that is not restricted to the mean-field states considered
in the results section. That is, it holds regardless of the form of
the approximate solution to the Schr\"odinger equation used.

The use of the dual representation in the context of
symmetry-projected techniques requires a numerical solution of the GHW
equations, as opposed to the direct representation where the
amplitudes are known a priori. The dual representation, however, can
still be useful in other contexts. As an example, one typically faces
numerical difficulties in evaluating expectation values of the state
$\hat{P}^q |\Phi \rangle$ whenever $|\langle \Phi | \hat{P}^q |\Phi
\rangle|$ is very small. In such a case, one can first rotate along a
dual representation to an intrinsic state $|\Phi_\chi \rangle$ chosen
such that it maximizes $|\langle \Phi_{\vartheta} | \hat{P}^q
|\Phi_{\vartheta} \rangle|$.

The existence of the dual representation of the GM also suggests
caution when trying to read significance into the structure of the
deformed states optimized in symmetry-projected methods: there are
multiple intrinsic states, with different deformation parameters, that
all lead to the same symmetry-projected wavefunction.

\section{Acknowledgments}

CAJH is grateful for support from a Wesleyan University start-up
package. The work at Rice University was supported by the
U.S. National Science Foundation under Grant No. CHE-1762320. GES is a
Welch Foundation Chair (Grant No. C-0036).

\appendix

\section{On the equivalence of the Goldstone manifold representations}
\label{sec:eqGM}

Consider a symmetry broken state $|\Phi \rangle$ and its expansion in
terms of normalized symmetry-adapted states $\{ \chi \}$:
\begin{equation}
  |\Phi \rangle = \sum_q |\chi_q \rangle \, c_q,
\end{equation}
with $c_q = \langle \chi_q | \Phi \rangle$. Here, the subscript $q$ on
$|\chi_q \rangle$ labels the symmetry of the states, such that
\begin{equation}
  \hat{Q} |\chi_q \rangle = q |\chi_q \rangle.
\end{equation}

As shown below, $\exp(i\theta \hat{Q}) \, |\Phi \rangle$ is expanded
in terms of the {\em same} set of symmetry-adapted states as $|\Phi
\rangle$:
\begin{equation}
  \exp(i\theta \hat{Q}) \, |\Phi \rangle = \sum_q \exp(i\theta
  \hat{Q}) \, |\chi_q \rangle \, c_q = \sum_q |\chi_q \rangle \, d_q
\end{equation}
with $d_q = \exp(i\theta q) \, c_q$.

A similar reasoning applies in the case of the dual
representation. The states $\exp(\vartheta \hat{Q}) \, |\Phi \rangle$
are also expanded in the {\em same} set of symmetry-adapted states as
$|\Phi \rangle$:
\begin{equation}
  \exp(\vartheta \hat{Q}) \, |\Phi \rangle = \sum_q \exp(\vartheta
  \hat{Q}) \, |\chi_q \rangle \, c_q = \sum_q |\chi_q \rangle \, b_q
\end{equation}
with $b_q = \exp(\vartheta q) \, c_q$. It follows that the direct and
dual representations are spanned by the same set of symmetry-adapted
states and are therefore equivalent.

As a corollary, note that
\begin{align}
  \hat{P}^q \exp(i\theta \hat{Q}) \, |\Phi \rangle &= \, |\chi_q \rangle
  \, c_q \, \exp(i\theta q), \\
  \hat{P}^q \exp(\vartheta \hat{Q}) \, |\Phi \rangle &= \, |\chi_q \rangle
  \, c_q \, \exp(\vartheta q).
\end{align}
That is, the same symmetry-projected states (up to arbitrary phase and
normalization factors) are obtained from the direct and dual
representations of the GM.

\section{Structure of the GCM kernels in $U(1)$ symmetry}
\label{sec:kernel}

In this section we provide a brief discussion of the structure of the
norm and Hamiltonian kernels associated with $U(1)$ symmetry
restoration in both the direct and dual representations.

In the direct representation, the overlap kernel $\mathcal{S}$ among
the states of Eq. \ref{eq:expiQ}, defined by
\begin{equation}
  \mathcal{S}(\theta',\theta) \equiv \langle \Phi_{\theta'} |
  \Phi_\theta \rangle,
\end{equation}
satisfies a translation symmetry
\begin{equation}
  \mathcal{S}(\theta',\theta) = \mathcal{S}(0,\theta-\theta').
\end{equation}
This, along with the periodicity of the $\theta$ domain, implies that
a Fourier transform of the norm kernel brings it to diagonal form. The
same holds true for the Hamiltonian kernel. In a discretized,
equi-spaced grid the norm kernel has a matrix structure of the form
\[
  \begin{pmatrix}
    a & b & c & d \\
    b & c & d & a \\
    c & d & a & b \\
    d & a & b & c \\
  \end{pmatrix}.
\]
That is, the norm (and Hamiltonian) matrix is circular and its
eigenvectors are given by a discrete Fourier transform.

In the dual representation, the overlap kernel $\mathcal{S}$ among
unnormalized states $|\tilde{\Phi}_\vartheta \rangle = \exp(\vartheta
\hat{Q}) \, |\Phi \rangle$, defined by
\begin{equation}
  \mathcal{S}(\vartheta',\vartheta) \equiv \langle
  \tilde{\Phi}_{\vartheta'} | \tilde{\Phi}_\vartheta \rangle,
\end{equation}
satisfies a translation symmetry
\begin{equation}
  \mathcal{S}(\vartheta',\vartheta) = \mathcal{S}(0,\vartheta+\vartheta').
\end{equation}
In this case, however, the domain of $\vartheta$ is not periodic. In a
discretized, equi-spaced grid the norm (and Hamiltonian) kernel has a
matrix structure of the form
\[
  \begin{pmatrix}
    a & b & c & d \\
    b & c & d & e \\
    c & d & e & f \\
    d & e & f & g \\
  \end{pmatrix}.
\]
The eigenvectors of such a matrix do not have an analytic
representation that is independent of the specific matrix
elements. Note, however, that the full matrix can be constructed from
evaluation of a linear (and not quadratic) number of terms.

\bibliography{paper}

\end{document}